\documentclass[sigconf]{acmart}

\usepackage{booktabs} 
\usepackage{subfigure}

\setcopyright{rightsretained}

\acmDOI{10.475/123_4}

\acmISBN{123-4567-24-567/08/06}

\begin{document}
\title{Collaborative Multi-modal deep learning for the personalized product retrieval in Facebook Marketplace}

\author{Lu Zheng}
\affiliation{%
  \institution{Facebook}
  \streetaddress{1 Hacker Way}
  \city{Menlo Park} 
  \state{California} 
  \postcode{94025}
}
\email{newluzheng@gmail.com}

\author{Zhao Tan}
\affiliation{%
  \institution{Facebook}
  \streetaddress{1 Hacker Way}
  \city{Menlo Park} 
  \state{California} 
  \postcode{94025}
}
\email{maximtzh1000@gmail.com}

\author{Kun Han}
\affiliation{%
  \institution{Facebook}
  \streetaddress{1 Hacker Way}
  \city{Menlo Park} 
  \state{California} 
  \postcode{94025}
}
\email{hankun@fb.com}

\author{Ren Mao}
\affiliation{%
  \institution{Facebook}
  \streetaddress{1 Hacker Way}
  \city{Menlo Park} 
  \state{California} 
  \postcode{94025}
}
\email{neroam@fb.com}
\renewcommand{\shortauthors}{Lu Zheng et al.}

\begin{abstract}
Facebook Marketplace \cite{marketplace} is quickly gaining momentum among consumers as a favored customer-to-customer (C2C) product trading platform.  The recommendation system behind it helps to significantly improve the user experience.  Building the recommendation system for Facebook Marketplace is challenging for two reasons: 1) Scalability: the number of products in Facebook Marketplace is huge. Tens of thousands of products need to be scored and recommended within a couple hundred milliseconds for millions of users every day; 2) Cold start: the life span of the C2C products is very short and the user activities on the products are sparse. Thus it is difficult to accumulate enough product level signals for recommendation and we are facing a significant cold start issue.  In this paper, we propose to address both the scalability and the cold-start issue by building a collaborative multi-modal deep learning based retrieval system where the compact embeddings for the users and the products are trained with the multi-modal content information. This system shows significant improvement over the benchmark in online and off-line experiments: In the online experiment, it increases the number of messages initiated by the buyer to the seller by +26.95\%; in the off-line experiment, it improves the prediction accuracy by +9.58\%.
\end{abstract}

%
%





\maketitle

\section{Introduction}
\label{intro}
\begin{figure}
\centering
\includegraphics[width=80mm]{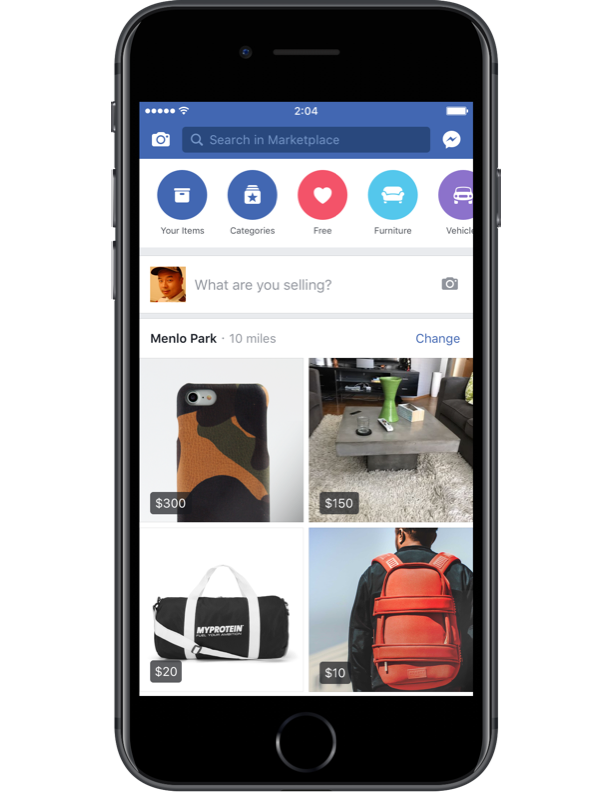}
\caption{Recommendation for the Facebook Marketplace.}
\label{fig_Marketplace}
\end{figure}

Facebook rolled out Marketplace where people can buy, sell and discover products \cite{marketplace} (Figure \ref{fig_Marketplace}). After the launch, Facebook Marketplace quickly obtain momentum  among the users as a favored C2C product trading platform. 

The Marketplace recommendation system is responsible for helping users to discover products that best fit their interests. Due to the large traffic,  the Marketplace recommendation system needs to process a huge number of requests every day. A common way of handling the mass recommendation in the industry is dividing the recommendation system into two cascade systems -- a retrieval system and a ranking system. The retrieval system is responsible for returning a short list from the candidate pool; The ranking system is responsible for giving the final recommendations out of the short list returned by the retrieval system. In the Marketplace recommendation system, we adopt this ranking-retrieval paradigm. 

In this paper, we will focus on how we use multi-modal deep learning to build the content based retrieval system. The retrieval stage recommendation is challenging for the following reasons:
\begin{itemize}
\item \textit{Real-time:} The retrieval system needs to serve millions of users to explore millions of the products in real time. The computation should be finished within a required latency (often on the order of O(100) milliseconds). Highly specialized and efficient serving system is needed for handling the Marketplace product retrieval.
\item \textit{Lack of the product taxonomy:} Most of the sellers on Facebook Marketplace are non-professional sellers. User research has shown that for these sellers, asking for too much product information will decrease their posting intent. Therefore, we only require the sellers to provide a coarse/broad category for the products when they post the products. This results in the lack of the accurate/fine grained product taxonomy for Facebook Marketplace and it is difficult to build a structured index for the products. 
\item \textit{Cold start and sparse user feedback:} Due to the huge inventory size and the buyers' preference of the newly posted products over the old products, the life spans of the C2C products are very short. A typical C2C product is active only for a few days before they fade out of Marketplace. It is difficult to collect enough product level signal within this short life span. 
\end{itemize}

However, there is a silver lining to the above challenges: in Marketplace, the content of the products, including the product images and the text descriptions, plays an important role in driving the buyer's purchase intent.  With that, we are inspired to build a retrieval system that extensively explores the relationship between the product content and the user activity. The core of this retrieval system is two collaboratively trained user and product deep neural network (DNN) models.  The models will digest the user profile and the product content to generate the user and the product representations. The product representations are used to index the products. At run time, the similarity between the user and the product representation is computed to retrieve the most relevant products for the user.

There are numerous works about recommendation systems. Existing recommendation systems can be roughly divided into three categories: collaborative filtering based approach\cite{collaborative_filering}, content based approach and the hybrid method approach. Collaborative filtering approach \cite{cf1, mf} makes use of the user activities or feedback without using the item content.  The content based approach make use of the user profile and the item description for recommendation \cite{Lang95newsweeder:learning}. The hybrid approach \cite{hybrid01,hybrid02,hybrid03} is a combination of the content based approach and the collaborative filtering based approach. 

Deep learning has successful applications in various areas such as computer vision (CV), image processing\cite{faster_rcnn} and the natural language processing (NLP) \cite{nlp_scratch}. Recently, there are also works applying deep learning in mass scale recommendation system \cite{youtube_dnn,deep_wide}. The recommendation is modeled as an extreme multi-class classification problem in \cite{youtube_dnn} and the user embeddings are learned based on the user's implicit feedback. 

While we also apply deep learning in the retrieval system, we focus on exploring the power of the deep learning model for the content understanding. As mentioned above,  the reasons for this are (1) in Marketplace, the users' purchase intent is driven by the content of the products. (2) the user feedback for the products is sparse due to the short life-span of the products. (3) The content information is relatively static compared with the history information. It is therefore easier to design the service system. 

This paper is organized as follows: a brief system overview is presented in Section \ref{system_overview}. Section \ref{sec_model} describes the retrieval system in details. In Section \ref{evaluation}, we provide the evaluation for the retrieval model. 

\section{System Overview}
\label{system_overview}
\begin{figure}
\centering
\includegraphics[width=80mm]{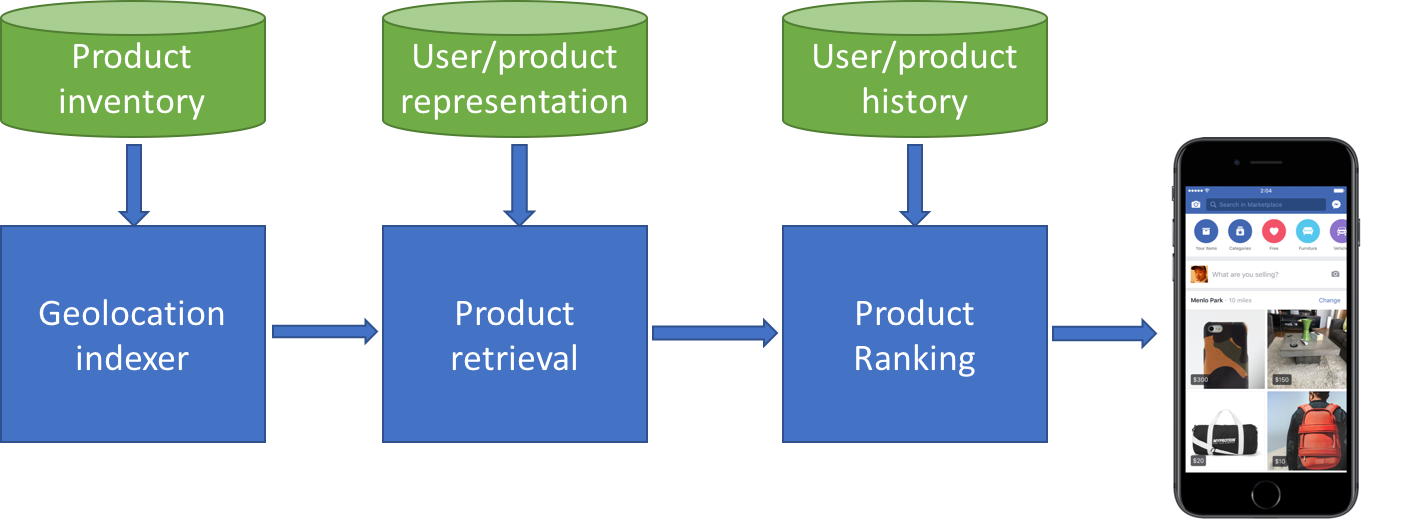}
\caption{Two stage recommender system for Marketplace}
\label{fig_retrieval_ranker}
\end{figure}

Facebook Marketplace is designed to enable users to discover and purchase nearby products. An overview of the recommendation system behind Marketplace is shown in Figure \ref{fig_retrieval_ranker}. The system consists of a geo-location indexer, a retrieval system and a ranking system. The geo-location indexer indexes all Marketplace products based on the product geo-location so that we can easily look up the nearby products for the users.  The retrieval system generates a list of the products from the local product pool. The ranking system then ranks the retrieved products to best match the context and the user interests.  

The retrieval system sits between the geo-location indexer and the ranking system. In a Marketplace recommendation request, a user and his/her interests are viewed as a query. The Marketplace retrieval system is designed to match the products with the queries based on the content of the products. 

\section{The Retrieval System Design}
\label{sec_model}
\begin {table}[H]
\caption{Notations}
\begin{center}
\begin{tabular}{|c|c|}
\toprule[1.5pt]
\hline
Notation & Description \\
\hline
$\mathcal{U}$ & User set \\
\hline
$\mathcal{I}_u$ & Product Item set available for the user $u$ \\
\hline
$i,j \in \mathcal{I}$ & product index \\
\hline
$u \in \mathcal{U}$ & user index \\
\hline
$\Theta$ & Model parameter \\
\hline
\bottomrule[1.25pt]
\end{tabular}
\end{center}
\end{table}

Given a user $u$ and a set of products that satisfy the this user's geo-location setting $\mathcal{I}_u$, the retrieval result $\mathcal{I}_u^*$ is set to be:
\begin{equation}
\mathcal{I}_u^* = Top\_N_{ i \in \mathcal{I}_u}(r_{ui}), 
\end{equation}
where $r_{ui}$ is the score between the user $u$ and the product $i$, $N$ is the total number of products retrieved.  

Since we need the serving system to be efficient, $r_{ui}$ is modeled as a dot product of a user "embedding" and a product "embedding", i.e., $r_{ui} = \phi(u) * \gamma(i)$ where $\phi$ and $\gamma$ represent the mapping from the user and the product to the latent embedding space. At serving time, we can simply perform a nearest neighbor search to obtain the top $N$ products for users. 

Now the problem is reduced to modeling the $\phi$ and $\gamma$. In this paper, $\phi$ and $\gamma$ will be collaboratively learned from the user and the product interaction. Given the run-time environment constraint, we want the models of $\phi$ and $\gamma$ to be as computationally light-weight as possible. Therefore, we choose to learn $\phi$ and $\gamma$ on relative static information to avoid the need for the frequent embedding update. For $\phi$, the input includes the \emph{user demographics} and the \emph{long term click history}; for $\gamma$, the input includes multiple modalities such as \emph{product images} and the \emph{product text}. We will discuss the details of how we model $\phi$ and $\gamma$ modeling in the following sections. 

\subsection{Collaborative deep learning with pairwise rank loss}
Since we want to learn the relative user preference for one item over another, we adopt \emph{pairwise collaborative filtering} to train the Marketplace retrieval models. Pairwise collaborative filtering with implicit user feedback is studied in \cite{collaborative_pairwisse}. Empirically, this approach is better than point-wise collaborative filtering \cite{one_class_cf}.  

The \textit{pairwise preference over the items} is defined as follows:
\begin{equation}
r_{ui} > r_{uj}, i,j \in \mathcal{I}_u,
\end{equation}
where the relationship $r_{ui} > r_{uj}$ means that the user $u$ prefer the item $i \in \mathcal{I}_u$ to the item $j \in \mathcal{I}_u$. 
In the ranking context, we relax the \textit{pairwise preference over items} to the \textit{pairwise preference over the item sets within a margin}\cite{collaborative_pairwisse}:
\begin{equation}
r_{u\mathcal{P_+}} - r_{u\mathcal{P_-}} > \xi, \mathcal{P_+} \subset \mathcal{I}_u, \mathcal{P_-} \subset \mathcal{I}_u,
\end{equation}
where $r_{u\mathcal{P_+}}$ and $r_{u\mathcal{P_-}}$ are the user $u$'s overall ranking scores on the items from the item-set $\mathcal{P_+}$ and $\mathcal{P_-}$, respectively. Here $\mathcal{P_+}$ and $\mathcal{P_-}$ depend on the user $u$ and should be represented as $\mathcal{P_+}(u)$ and $\mathcal{P_-}(u)$. For simplicity we drop this dependency in the equation for the rest of this paper. $\xi$ is the margin, which is set to $1$ in this paper.

In our work, we will combine the deep learning model with pairwise collaborative filtering to learn $\phi$ and $\gamma$. 

\begin{figure*}
\centering
\includegraphics[width=180mm]{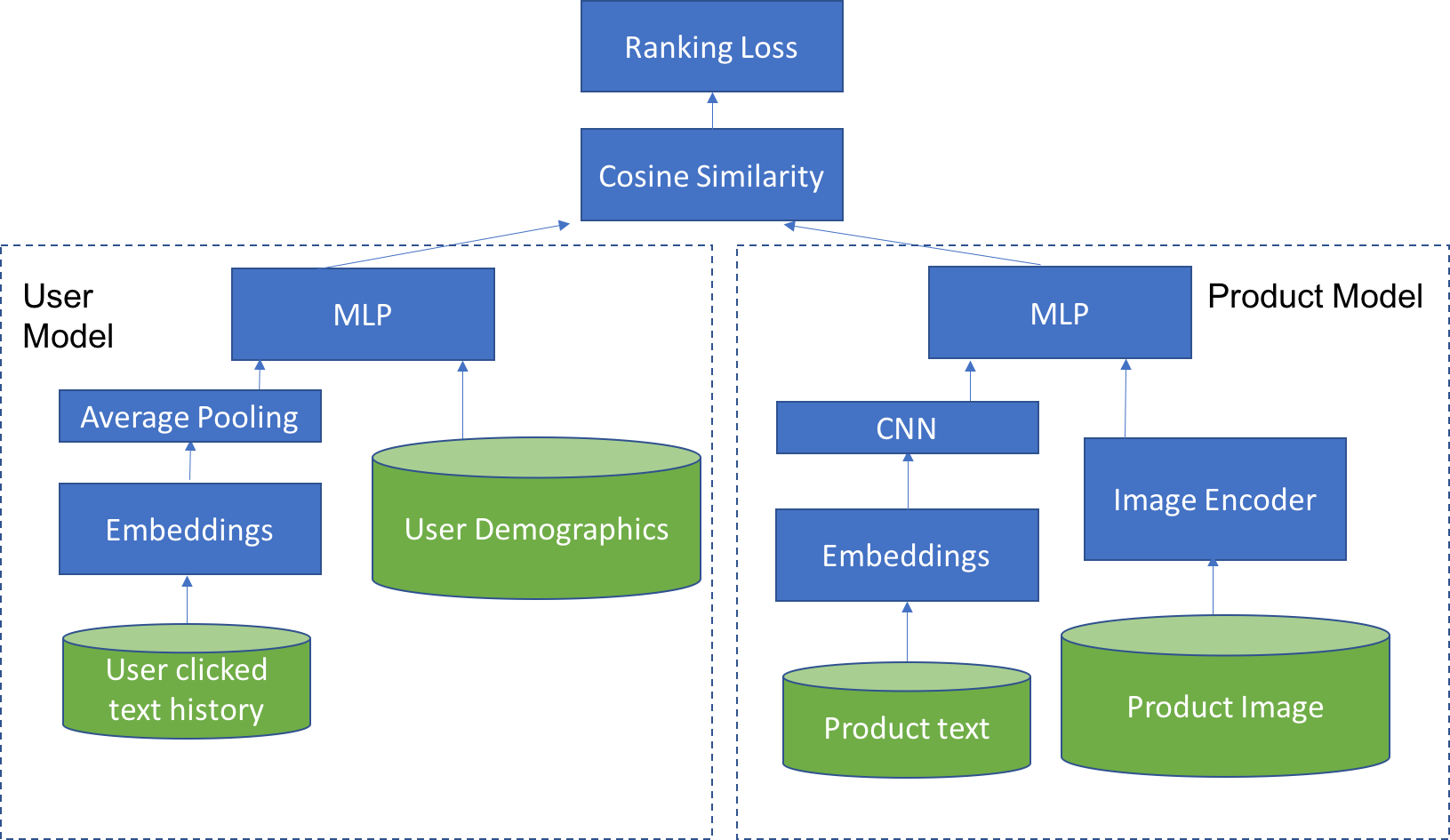}
\caption{Deep Neural Network structure for User and Product}
\label{fig_pairnn}
\end{figure*}

The overall model is shown in Figure \ref{fig_pairnn}. The model consists of two parts: a user model $\phi(u)$ and a product model $\gamma(i)$. They are jointly trained on the events logging of the products: We label an impression of the product to be positive or negative based on whether the impression led to any messages sent from the buyer to the seller. We use the pairwise rank loss to guide the training of the model.  We call the model pairNN for short. 
\subsubsection{pairwise loss function}


Let $\Theta$ be the model parameter. The objective function for the training is:
\begin{equation}\label{objectivefunction}
\min_{\Theta} \sum_{u \in \mathcal{U}} \sum_{p_+ \in \mathcal{P_+}}  \sum_ {p_- \in \mathcal{P_-}} \max(0, r_{u{p_-}} - r_{u{p_+}} + \xi),
\end{equation}

For each user $u$, we consider the product $p$ as a positive example $p_+$ if this user initiate a message thread with the seller about this product and a negative example $p_-$ otherwise. Therefore, a training example is a tuple of $(u, p_+, p_-)$.

In practice, with the data from all the users, we collaboratively train the model by optimizing the following objective function: 
\begin{equation}\label{rankingloss}
\min_{\Theta} \sum_{(u, p_+, p_-) \in \mathcal{E}} \max(0, r_{u{p_-}} - r_{u{p_+}} + \xi),
\end{equation}
where $\mathcal{E}$ presents all the training samples. 


In order to compute the ranking scores $r_{u{p_+}}$ and $r_{u{p_-}}$, we use deep neural networks (DNNs) to learn the embeddings for the user and the product separately, and then use the cosine similarity between the embeddings as the ranking score. 

\subsubsection{User model}
We design the user model to take the heterogeneous information including the user keywords and the user demographic features.

In the modeling, we use an embedding layer to map each keyword to a vector. The embedding is initialized with a set of pre-trained word2vec embeddings \cite{mikolov2013distributed} and updated during the training. An average pooling layer is used to aggregate the embedding vectors of all keywords. The output of the averaged word embeddings  
 is concatenated with the user demographic information in the form of the dense feature, and then sent into a multilayer perceptron (MLP) to learn the user embedding vector. 
 
 \subsubsection{Product model}
The features of the product model include the text and the images of the products. The text feature is the word sequence of the product title and description. To model the word sequence, we first map each word in the word sequence to an embedding and then feed them to the convolutional neural networks (CNNs) \cite{collobert2011natural,kim2014convolutional}. The word sequence is then converted into a fixed length text embedding. For image feature, we utilize a Facebook internal image classification model to encode the images of the products. The image encoder is a 50-layer ResNet \cite{resnet} pre-trained on the image classification task and stays fixed in the training procedure. The output of the image encoder and the text embedding are concatenated and sent into a multilayer perceptron (MLP). 

With the vector representations of both the user and the product, we compute the cosine similarity as the final score $r_{ui}$. At training time, we optimize the similarity score with a ranking loss defined in Eq. \ref{rankingloss}.

\section{Experiments and Results}
\label{evaluation}
\subsection{Training Data}
We generate the training data from the user message log through Facebook Marketplace. The data we collected is limited and anonymized. The content of the message is unaccessible to the researchers -- we only log whether a message is initiated by a buyer to a seller to label whether the product is a positive or negative example for this buyer. A total of 5+M messages are sampled from a couple of weeks message log as the positive examples. An additional 70+M impressions are sampled as the negative examples. Since we are using the pair-wise loss function in the training, the required training sample is a tuple $(u, p_+, p_-)$, where $u \in \mathcal{U}$, $p_+ \in \mathcal{P}_+$ and $p_- \in \mathcal{P}_-$. Thus we pair up one positive example and one negative example for a given user and form the pair-wise training data. 


\subsection{Experiment Setup}
We evaluate the models with both the online and the offline experiment.

The online experiment evaluates the performance of the embeddings in the online retrieval system for Marketplace. In the retrieval task, for user $u$, we will first fetch the most recent $M$ products from the set of $\mathcal{I}_u$ that satisfies the user's location and radius setting and then select the top $N$ products based on the score $r_{ui}$ out of the $M$ products. The selected $N$ products are then used as the input to the ranking system.

The results are evaluated by measuring the number of messages initiated by the buyers in the A/B test groups. The system randomly choose de-identified users on Facebook Marketplace platform and split them into the test and the control groups for A/B test. For different groups we are using different approaches to retrieve product candidates. The ranking stage is the same across different groups in the experiment though.

The offline experiment evaluates the model in a classification setting: the evaluation data is generated the same way as the training data where we pair up the positive $p_+$ and negative samples $p_-$ for a given user $u$. For an evaluation sample $(u, p_{+}, p_{-})$,  we call the classification correct if $r_{up_{+}} > r_{up_{-}}$ and wrong otherwise. 

\subsection{Metrics}
In the online experiment, we measure the performance of the model based on the message initiation by the buyers. 
    \begin{itemize}
    \item \textbf{Message initiation:} We count the number of initial messages sent from buyers to sellers through Facebook Marketplace. 
    \end{itemize}
   
 In the offline experiment, the model is measured by the \textbf{accuracy} and the \textbf{average loss} defined as follows. 
\begin{itemize}
\item \textbf{Accuracy:} Considering a binary classifier on the dataset consists of tuple $(u, p_+, p_-)$, the accuracy of models is defined as, $Acc =  \sum_{(u, p_+, p_-)\in\mathcal{T}}\delta(r_{up_+}-r_{up_-}) / |\mathcal{T}|$, where $\delta(x) = 1$ if $x > 0$ and $\delta(x)=0$ otherwise. Thus $\sum_{(u, p_+, p_-)\in\mathcal{T}}\delta(r_{up_+}-r_{up_-})$ denotes the number of tuple samples in the test set that have higher score for positive product than negative one. With higher accuracy, the model in general can retrieve more positive products then negative ones.

\item \textbf{Average loss}   The average loss is defined as $\textbf{average loss} = \sum_{(u, p_+, p_-)\in\mathcal{T}}\max(0, r_{u{p_-}} - r_{u{p_+}} + \xi) / |\mathcal{T}|$. The average loss measures how well the model can distinguish the positive and the negative samples. 
\end{itemize}

\subsection{Baseline}
\label{sec:exp_baseline}
We compare the pairNN based retrieval with two other retrieval approaches namely, time-based retrieval and word2vec-based retrieval.
\begin{itemize}
\item \textbf{Time-based Retrieval}: The time-based retrieval is the most nature way of retrieving the products. In this approach, the most recent $M$ products are selected and fed to the ranking system. 
\item \textbf{Word2vec-based Retrieval}: In this approach, we fetch the products based on the cosine similarity between the user keywords and the product title keywords.  The user keywords are again collected from the posts that the user clicked. We first train the word2vec keyword embeddings on a corpus of the product title and description. Given a list of user keywords $(k_1, ... k_m)$ and a list of product title keywords $(t_1, ... t_n)$, we map each of the user and product title keywords to the word2vec embedding $(v_{k_1}, ... v_{k_m})$ and $(v_{t_1}, ..., v_{t_n})$.  The text similarity is computed as follows: 

\begin{equation}
similarity(user, product) =  \bar{v}_k * \bar{v}_t
\end{equation}
where $\bar{v}_k = \sum_{i = 1}^m v_{k_i} / m $ and $\bar{v}_t = \sum_{i = 1}^n v_{t_i} /n $. We retrieve the top $N$ products based on the similarity score. 
\end{itemize}


\subsection{Summary of the experiment result}
In the offline experiment, We compare the \textbf{accuracy} and the \textbf{average loss} of the pairNN model and the word2vec-based model. The pairNN model is trained on the 10+ days of Facebook Marketplace message data and the world2vec model is trained on a Marketplace product description corpus. 

In pairNN, the MLP blocks in the user model and the product model both have three layers: the input layer has 100 perceptrons; the hidden layer has 100 perceptrons and the output layer has 50 perceptrons.  On the product side, the image encoder is a 50 layer pre-trained ResNet.  

The results are summarized in Table \ref{prediction_performance}. From the offline results, we can see that:
\begin{itemize}
\item Given the text only based approach, pairNN has better accuracy and average loss. This matches our expectation since the pairNN is trained supervisedly on the product event data while the word2vec embeddings are trained unsupervisedly. 
\item We compare the performance of pairNNs trained with different product content as the input: a) the text including the product title and description. b) the user uploaded image for this product c) the text and the image. From Table \ref{prediction_performance} we can see the multi-modal model generates the best accuracy.  
\end{itemize}

\begin {table}[H]
\caption{Offline result}
\label{prediction_performance}
\begin{center}
\begin{tabular}{c|c|c}
\toprule[1.5pt]
\hline
  & Accuracy & Average Loss \\
\hline
\hline
Word2vec-based  & 0.6312 & 0.8837 \\
\hline
\hline
\multicolumn{3}{c}{} \\
\hline
\hline
PairNN &Accuracy & Average Loss \\
\hline 
\hline
Text only & 0.6656 & 0.7598 \\
Image only &0.6621 & 0.7650 \\
\hline
\textbf{Text + Image} & \textbf{0.6917} & \textbf{0.7239} \\
\hline
\bottomrule[1.25pt]
\end{tabular}
\end{center}
\end{table}



\begin{figure}
\centering
\subfigure[Loss curve]{
\includegraphics[width=80mm]{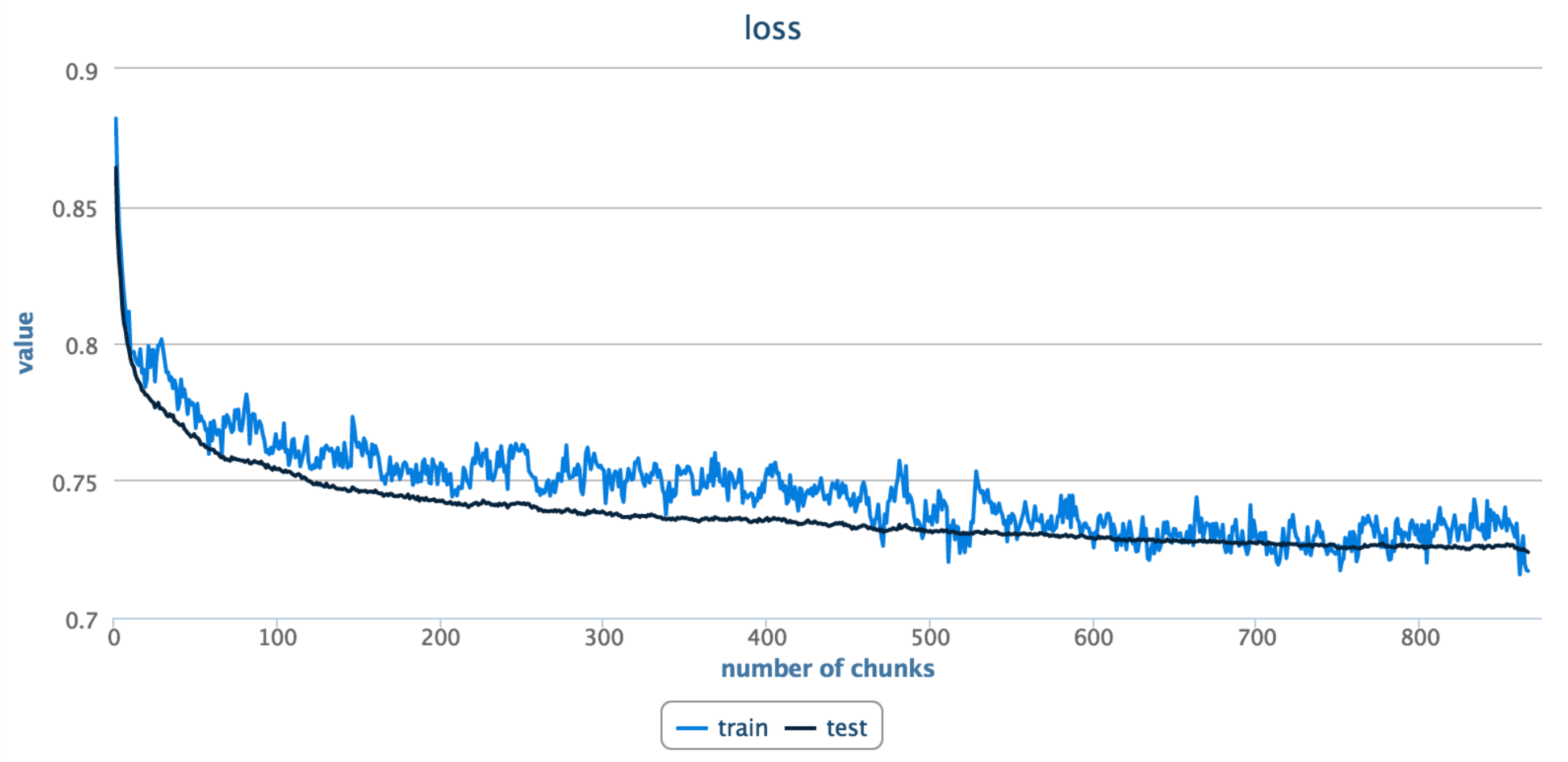}
}
\subfigure[Accuracy curve]{
\includegraphics[width=80mm]{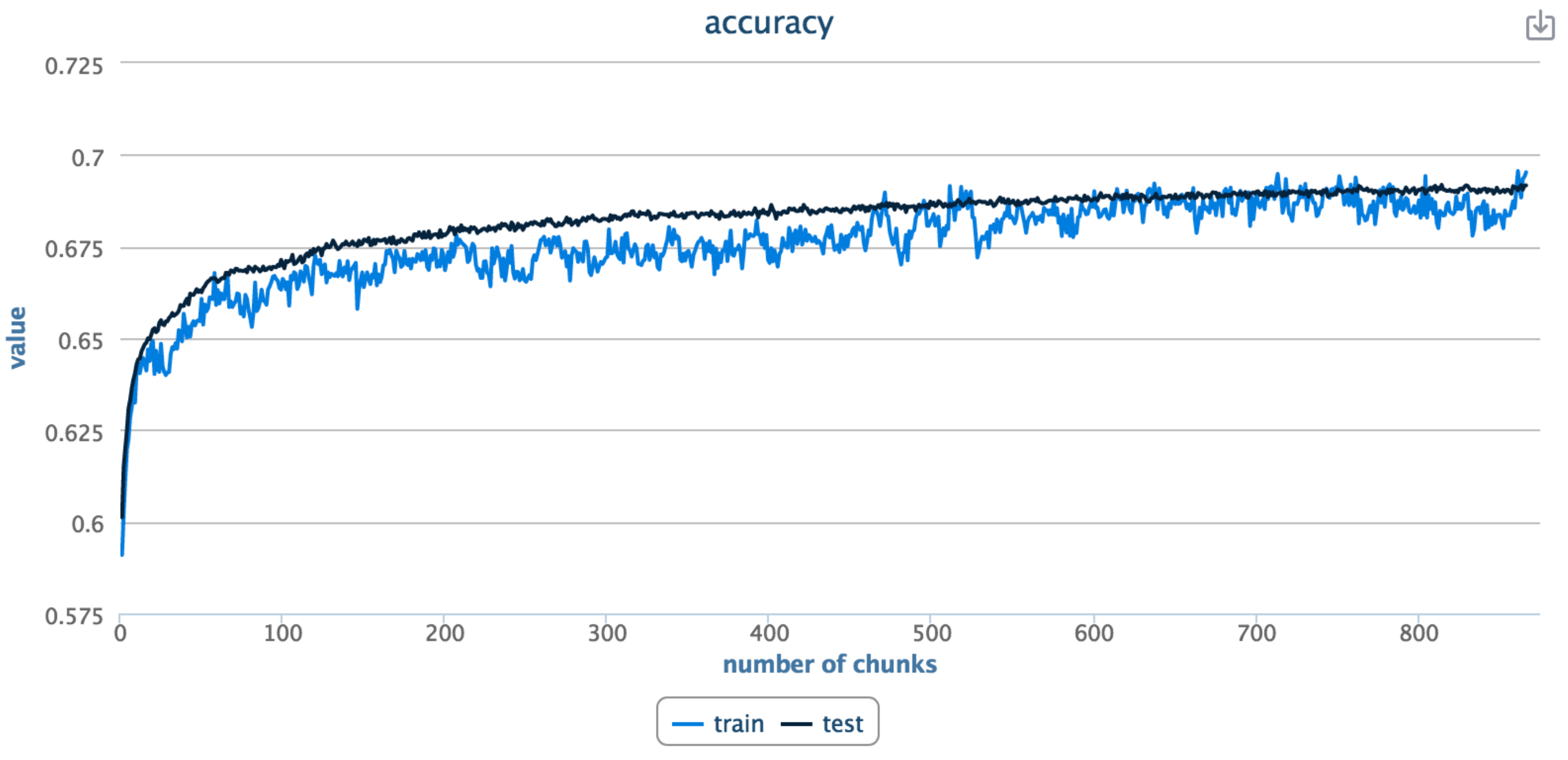}
}
\subfigure[Distribution of prediction score in test set]{
\includegraphics[width=80mm]{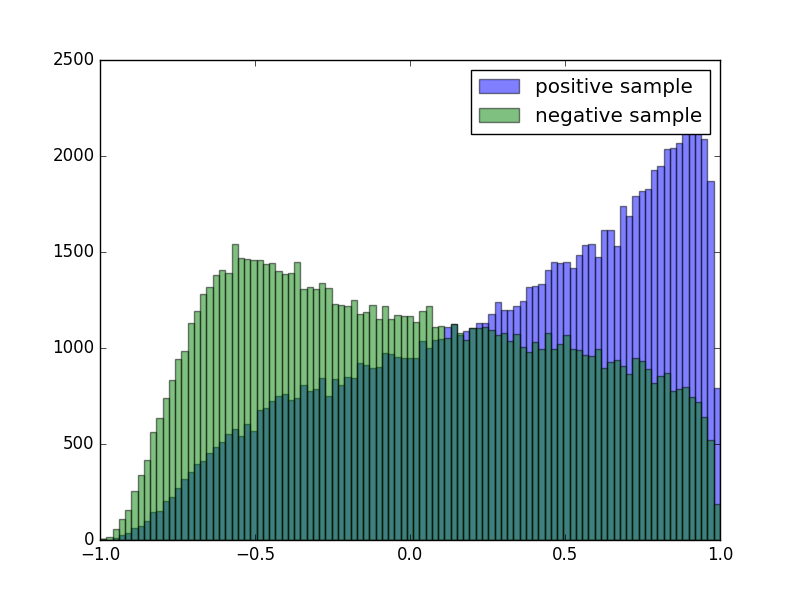}
}
\caption{Learning curve and Distribution of prediction score}
\label{fig_loss_perf}
\end{figure}

The learning curves and distribution of prediction scores of the multi-modal model on the test dataset are shown in Fig. \ref{fig_loss_perf}. 

In the online A/B experiment, we compare the multi-modal pairNN based retrieval system, the word2vec-based retrieval system and the time based retrieval.  The word2vec-based retrieval and time based retrieval are described in Section \ref{sec:exp_baseline}.  

We measure the quality of the retrieval results by the number of messages initiated by the users. The online experiment ran for two weeks and Table \ref{online_perf} summarizes the experiment result. Compared with the time based retrieval, word2vec based retrieval increases the number of messages by 10.39\% and the pairNN based retrieval increases the number of messages by 26.95\%. 


\begin {table}[H]
\caption{Online Performance}
\label{online_perf}
\begin{center}
\begin{tabular}{c|c}
\toprule[1.5pt]
\hline
Methods & \#message \\
\hline
Time-based (baseline) & 308k\\
\hline
Word2vec-based  & 340k  ($\Uparrow10.39\%$) \\
\hline
Proposed & 391k ($\Uparrow26.95\%$) \\
\hline
\bottomrule[1.25pt]
\end{tabular}
\end{center}
\end{table}
\section{Conclusion and future work}
In this paper, we show the model architecture and the system design for Facebook Marketplace product retrieval. Given the real-time constraint and the large traffic volume, we need to extensively explore content information for the retrieval. In addition to that, the content based retrieval system can also help us resolve the cold start problem. The proposed pairNN model enable us to leverage the multi-modal information from products and shows significant improvement for the retrieval quality in both online and offline evaluation. 


\bibliographystyle{ACM-Reference-Format}
\bibliography{sigproc} 

\end{document}